\documentclass[conference]{IEEEtran}
\IEEEoverridecommandlockouts

\usepackage{cite}
\usepackage{threeparttable}
\usepackage{amsmath,amssymb,amsfonts}
\usepackage{algorithmic}
\usepackage{graphicx}
\usepackage{textcomp}
\usepackage{xcolor}
\usepackage{comment}
\usepackage{booktabs}
\usepackage{multirow}
\usepackage{subcaption}
\usepackage[hidelinks]{hyperref}
\usepackage{makecell}
\usepackage{tikz}

\newcommand{\sep}[1][]{\,/\,}

\def\BibTeX{{\rm B\kern-.05em{\sc i\kern-.025em b}\kern-.08em
    T\kern-.1667em\lower.7ex\hbox{E}\kern-.125emX}}

\newcommand\copyrighttext{%
  \footnotesize
  \textcopyright~2026 IEEE. Personal use of this material is permitted.
Permission from IEEE must be obtained for all other uses, in any current or
future media, including reprinting/republishing this material for advertising
or promotional purposes, creating new collective works, for resale or
redistribution to servers or lists, or reuse of any copyrighted component
of this work in other works.
}

\newcommand\copyrightnotice{%
\begin{tikzpicture}[remember picture,overlay]
\node[anchor=south,yshift=16pt] at (current page.south)
  {\fbox{%
    \parbox{\dimexpr\textwidth-\fboxsep-\fboxrule\relax}{%
      \copyrighttext
    }%
  }};
\end{tikzpicture}%
}

\title{Spooftral: Can Voxtral Audio-Language Model Detect Speech Spoofing?}

\author{
\IEEEauthorblockN{
Avishai Weizman\textsuperscript{1} \qquad
Yehuda Ben-Shimol\textsuperscript{1} \qquad
Itshak Lapidot\textsuperscript{2,3}
}

\IEEEauthorblockA{
\textsuperscript{1}School of Electrical and Computer Engineering,
Ben-Gurion University of the Negev, Israel \\
\textsuperscript{2}Department of Electrical Engineering,
Afeka the Academic College of Engineering, Israel \\
\textsuperscript{3}Avignon University, LIA, France
}
}

\begin{document}

\maketitle

\copyrightnotice

\begingroup
\renewcommand{\thefootnote}{}
\renewcommand{\footnoterule}{}

\makeatletter
\def\@makefntext#1{#1}
\makeatother

\footnotetext{\scriptsize
GitHub repository: \url{https://github.com/avishai111/Spooftral}
}

\endgroup

\begin{abstract}
\textit{Self-supervised learning} (SSL) \textit{countermeasures} (CMs) have shown strong performance in recent years. However, they often show degraded performance while facing unseen spoofing attacks and mismatched conditions. This study examines the Voxtral \textit{audio-language model} (ALM) framework for spoofing detection, as a step toward combining CM capabilities within the ALM framework. We analyze how Voxtral captures spoofing cues through audio-text processing and propose an instruction-guided approach that uses label-sequence likelihoods to evaluate bonafide and spoofed speech. Experiments on the ASVspoof databases show that without task-specific adaptation, the LLM layers emphasize semantic representations, reducing the separability of spoof-discriminative acoustic cues compared to the Whisper-based audio encoder. Consequently, spoofing-related information becomes less separable after language-model processing. We also applied lightweight adaptation using weight-\textit{decomposed low-rank adaptation} (DoRA) to the Voxtral model and propose the Spooftral model, achieving an \textit{equal error rate} (EER) of 4.25\% on the ASVspoof5 evaluation set.
\end{abstract}


\section{Introduction}

\textit{Automatic speaker verification} (ASV) systems are widely used in modern biometric applications, including access control, financial authentication, and voice interfaces~\cite{mittal2022automatic,naika2018overview}. Despite their rapid adoption, ASV systems remain vulnerable to spoofing attacks, such as replay, \textit{text-to-speech} (TTS), and \textit{voice conversion} (VC), which can compromise system reliability. As a result, developing effective \textit{countermeasures} (CMs) for spoofing detection is a crucial research direction in recent years~\cite{kamel2025surveythreatsvoiceauthentication,almutairi2022review}.

Several databases have been released to support research on spoofing detection, including the ASVspoof databases~\cite{wu15e_interspeech,kinnunen2017asvspoof,wang2020asvspoof,10155166,WANG2026101825}. These databases introduce a variety of attack types, acoustic conditions, and recording environments, providing benchmarks for evaluating CM robustness. Despite these advances, CM systems still suffer from the generalization problem to unseen spoofing attacks~\cite{muller2022does,zhang2023impact,pascu24_interspeech,weizman2024tandem,weizmanCSCML,weizman2026spoofing}. Recent findings~\cite{weizman25_interspeech} further indicate that the ASVspoof5 database~\cite{WANG2026101825} introduces not only more challenging attacks, but also a notable shift in the distribution of bonafide speech across its sets compared to the ASVspoof2019 database~\cite{wang2020asvspoof}, highlighting the growing need for improved spoofing detection approaches.

Today, \textit{self-supervised learning} (SSL) approaches dominate the spoofing detection landscape~\cite{serrano2025improving,tak2022automatic,kulkarni24_asvspoof,tran25b_interspeech}. Models such as Wav2Vec2~\cite{baevski2020wav2vec}, HuBERT~\cite{Hsu2021HuBERTSS}, and WavLM~\cite{WavLM} leverage large-scale pre-training on unlabeled audio to learn speech representations and have achieved strong results in recent ASVspoof challenges~\cite{wang24_asvspoof}. In parallel, \textit{weakly supervised learning} models, such as Whisper~\cite{Radford2022RobustSR}, trained on massive audio-text corpora, have demonstrated strong transferability across a wide range of speech tasks~\cite{kawa23b_interspeech,ozyilmaz25_interspeech,liang25c_interspeech}. However, despite its strong performance, Whisper has shown limited performance for spoofing detection tasks. In existing studies~\cite{10.1007/978-3-031-61382-1_8,kawa23b_interspeech}, Whisper-based representations improve detection performance but degrade under challenging conditions such as the ASVspoof2021 \textit{deepfake} (DF) challenge. This raises an open question about how well these representations are suited for the spoofing detection tasks that rely on fine-grained acoustic cues (e.g., spectral and temporal details). 

More recently,~\textit{audio-language models} (ALMs) have emerged as an alternative to acoustic formulations by jointly modeling speech signals and textual information within a unified framework. These models are characterized by larger parameter counts and are trained on massive, multi-modal databases pairing audio and text. Such pre-training provides a strong initialization for fine-tuning across downstream speech semantic tasks (i.e., tasks requiring an understanding of spoken content), including \textit{automatic speech recognition} (ASR) and \textit{speech-to-speech translation}, and in some cases, improves interpretability through text-based reasoning~\cite{ma2025audio}.

This work represents a step toward incorporating spoofing detection into unified ALM frameworks that can perform multiple speech tasks within a single model~\cite{zhang2023speechgpt,tang2023salmonn,ghosh2025audio,Qwen-Audio}, with the long-term goal of enabling both speech understanding and spoofing detection within a common audio-language framework, without any use of an external CM system. The main contribution is an analysis of how spoof-discriminative cues propagate within the Voxtral ALM, showing that spoofing detection requires task-specific training. To enable this analysis, we formulate spoofing detection as an instruction-guided ALM task~\cite{Qwen-Audio,tang2023salmonn}, rather than a conventional acoustic classification problem, and adapt Voxtral~\cite{liu2025voxtral} to the spoofing detection task using a lightweight training method (e.g., weight-\textit{decomposed low-rank adaptation} (DoRA)~\cite{liu2024dora}). In addition, we provide, to the best of our knowledge, one of the first evaluations of the Whisper-based audio encoder in the Voxtral model on the ASVspoof databases and compare embeddings extracted before and after the frozen \textit{large-language model} (LLM) layers in the Voxtral ALM framework.

The remainder of this paper is organized as follows. \autoref{sec:Related_Work} reviews related work on ALMs for the spoofing detection task and the Voxtral architecture; \autoref{sec:Proposed_System} presents the proposed generative label-likelihood classification method; \autoref{sec:databases_Evaluation_Metrics} describes the databases and evaluation metrics, and the experimental results are shown in \autoref{sec:Experiments_and_Results}. Finally, \autoref{sec:Conclusions_and_Discussion} concludes the paper and outlines the directions for future research.

\section{Related Work} \label{sec:Related_Work}
Research on ALMs has grown with the emergence of large multi-modal architectures capable of processing speech and text within a shared representation space~\cite{su2025audio,peng2025survey,deshmukh2023pengi}. Modern systems employ dedicated audio and language encoders, sometimes augmented with alignment modules or pre-trained language models to enhance cross-modal integration.  In many ALMs, the Whisper encoder models~\cite{Radford2022RobustSR} have become a common choice for the audio encoder due to their strong robustness~\cite{ghosh2025audio,Qwen-Audio,Tian2025UALMUA,shu2023llasm,li2024whisma}.
Following~\cite{su2025audio}, ALMs can be categorized into four main groups. Two-Tower models use separate encoders whose outputs are mapped into a common embedding space, such as the \textit{Contrastive Language-Audio pre-training} (CLAP) model~\cite{wu2023large}. Two-Head architectures rely on a single encoder with modality-specific projectors and a language model operating on top, as in SpeechGPT~\cite{zhang2023speechgpt}, Pengi~\cite{deshmukh2023pengi}, SALMONN~\cite{tang2023salmonn}, Audio Flamingo~\cite{ghosh2025audio}, and Qwen-Audio~\cite{Qwen-Audio}. One-Head designs construct a unified multi-modal input and use a single encoder to jointly process both modalities~\cite{han2023onellm,su2025audio}. The fourth category includes cooperative frameworks that employ an LLM as planning agent to enable flexible multi-modal interaction~\cite{li2024survey}. 
These architectural choices influence how acoustic and linguistic information is fused and represented within the model.

\begin{figure}[t]
    \centering
    \includegraphics[width=1\linewidth]
    {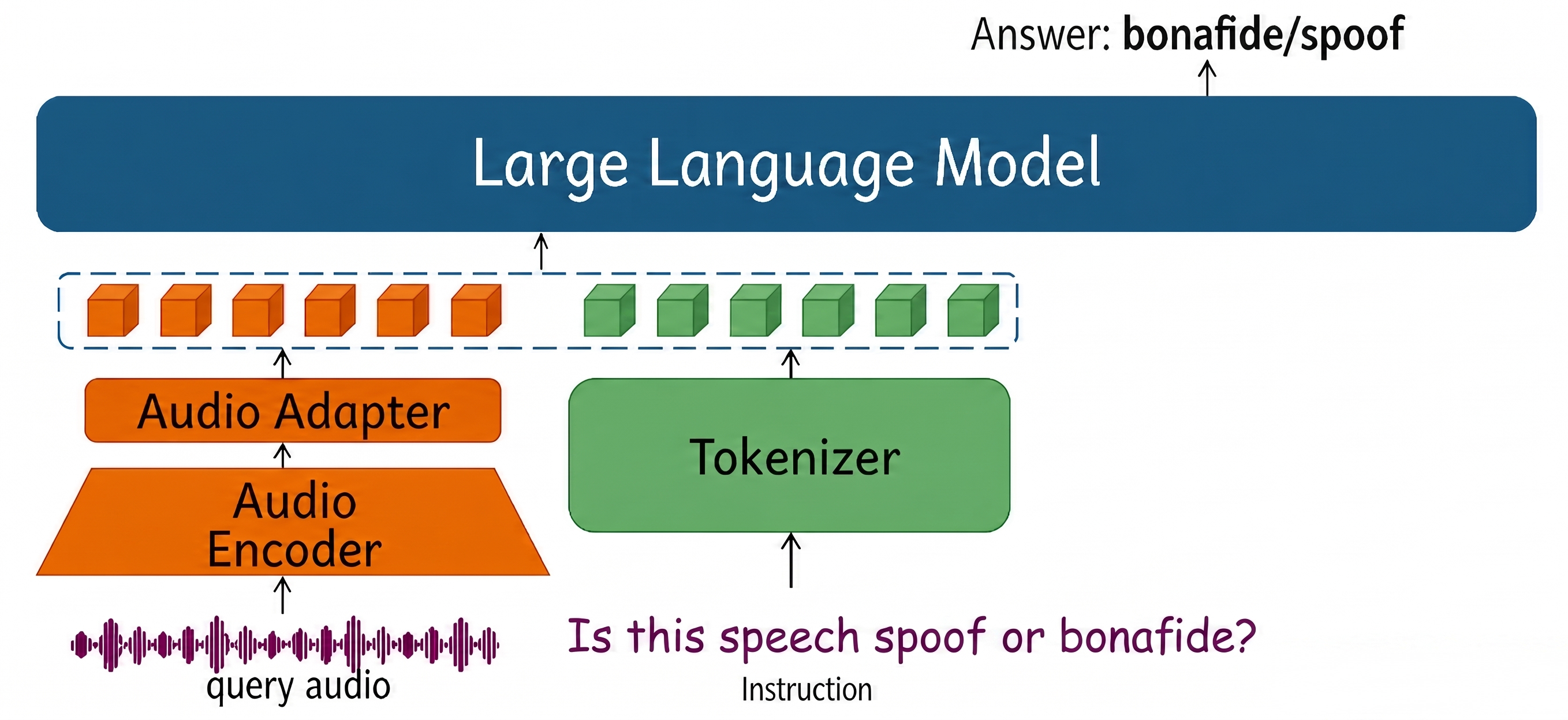}
    \caption{Overview of the Spooftral model based on Voxtral architecture, where audio and text inputs are jointly processed and spoofing decisions are obtained with likelihood scoring.}
    \label{fig:architecture}
    \vspace{-4mm}
\end{figure}

The Voxtral ALM model in~\cite{liu2025voxtral} belongs to the Two-Head category and is illustrated in~\autoref{fig:architecture}. Voxtral employs a dedicated audio encoder based on the Whisper Large-v3 model~\cite{Radford2022RobustSR}, followed by a temporal downsampling stage (audio adapter layers) composed of linear layers and an activation function, reducing the sequence length before projection into the embedding space of a pre-trained LLM. Voxtral is trained in three stages: pre-training, supervised fine-tuning, and preference alignment (e.g., \textit{Direct Preference Optimization} (DPO)~\cite{rafailov2023direct,guo2024direct}), which cover a wide range of speech understanding and reasoning tasks, and enable strong performance across diverse audio processing tasks~\cite{liu2026voxtralrealtime,liu2025voxtral}.

Recent studies on ALM-based CM systems highlight both the promise and limitations of this method. The work in~\cite{dutta25b_interspeech} reports that ALM-based CM systems are sensitive to quantization, with performance degradation when moving to lower-precision formats. In parallel, the study in~\cite{gu2025allm4add} presents an evaluation of ALMs for spoofing detection using Qwen-Audio~\cite{Qwen-Audio}, showing strong performance across databases, including ASVspoof2019~\cite{wang2020asvspoof} and the In-the-Wild database~\cite{muller2022does}. However, existing ALM-based CM studies mainly focus on end-to-end detection performance, with limited investigation of how spoof-discriminative information propagates across model stages. This motivates an evaluation of Voxtral-based representations, comparing spoof-discriminative information at the audio-adapter output, after the frozen LLM layers, and after task-specific adaptation.

\section{Generative Label-Likelihood Classification for Spoofing Detection}
\label{sec:Proposed_System}

This section reformulates spoofing detection as a generative label-likelihood classification framework, enabling instruction-guided modeling while evaluating the system using the \textit{equal error rate} (EER), a common metric for assessing CM systems, which allows comparison with other CM approaches. This formulation is compatible with unified ALM frameworks in which spoofing detection can be used together with other speech tasks that may require free-form generation, such as speech question answering. However, unlike these generative tasks, spoofing detection requires a controlled and reproducible decision. Therefore, for this task, we use a fixed inference prompt together with a fixed set of label tokens.

Given an audio input $a$ and a prompt $p$, Voxtral defines a conditional distribution $P(\cdot \mid a,p;\theta)$ over the token sequences, where $\theta$ denotes the model parameters. Although this distribution is defined over the entire vocabulary $\mathcal{V}$, no sampling is performed for the spoofing decision. Instead, the scores are computed from the raw model logits over the predefined label sequences and without applying temperature scaling in the spoofing-detection path. The final decision is obtained by computing length-normalized label log-likelihoods for ``\textit{bonafide}'' and ``\textit{spoof}'' and using the difference in their scores as the detection score.

We define a target vocabulary $\mathcal{V}_{\text{tar}} \subset \mathcal{V}$ containing the tokens used to form the class label strings. Each class label is represented by a sequence of target tokens $Y = (t_1, \ldots, t_{|Y|})$, where $t_i \in \mathcal{V}_{\text{tar}}$ and $|Y|$ denotes the length of the sequence. We use two label token sequences, $Y_{\mathrm{bf}}$ and $Y_{\mathrm{sp}}$, corresponding to the strings ``\textit{bonafide}'' and ``\textit{spoof}''. We intentionally use meaningful label tokens rather than arbitrary symbols (e.g., ``0/1''), which may allow the model to leverage its pre-trained semantic representations to associate acoustic cues with semantic labels. However, the likelihood-scoring framework is not restricted to the prompt-label interface used in this work. Other fixed interfaces can be defined by choosing a different but fixed prompt and corresponding fixed label tokens, provided that the same interface is used consistently during training and inference. To avoid potential length bias, we normalize the log-likelihood of each label-sequence by its token length.
The label-sequence score is defined as the length-normalized log-likelihood of the token sequence:
\begin{equation}
\frac{1}{|Y|}\log P(Y \mid a, p; \theta) =
\frac{1}{|Y|}
\sum_{i=1}^{|Y|}
\log P\left(t_i \mid a, p, t_{1:i-1}; \theta\right)
\end{equation}
where $t_{1:i-1} = (t_1, \ldots, t_{i-1})$ denotes the generated tokens and $|Y|$ is the number of tokens in sequence $Y$.
To obtain a binary decision (\textit{bonafide} vs.\ \textit{spoof}), we evaluate two label token sequences, $Y_{\mathrm{bf}}$ and $Y_{\mathrm{sp}}$, corresponding to the bonafide and spoof hypotheses, respectively, i.e., \textit{bonafide} (tokens \texttt{bon}, \texttt{af}, \texttt{ide}) and \textit{spoof} (tokens \texttt{sp}, \texttt{o}, \texttt{of}). Their length-normalized log-likelihoods are denoted by $\ell_{\mathrm{bf}}=\frac{1}{|Y_{\mathrm{bf}}|}\log P(Y_{\mathrm{bf}}\mid a,p;\theta)$ and $\ell_{\mathrm{sp}}=\frac{1}{|Y_{\mathrm{sp}}|}\log P(Y_{\mathrm{sp}}\mid a,p;\theta)$.
A detection score is obtained from the normalized log-likelihood difference between the spoof and bonafide hypotheses:
\begin{equation}
S = \ell_{\mathrm{sp}} - \ell_{\mathrm{bf}} 
\end{equation}
The score $S$ serves as the detection score used to compute the EER at inference. For training, the model optimizes a class-weighted \textit{cross-entropy} loss over the two label-sequence scores.
\begin{equation}
\mathcal{L}_{\mathrm{CE}}
= - w_y \log
\frac{\exp(\ell_y)}
{\exp(\ell_{\mathrm{bf}})+\exp(\ell_{\mathrm{sp}})}
\end{equation}
where $\ell_{\mathrm{bf}}$ and $\ell_{\mathrm{sp}}$ denote the normalized log-likelihoods of the \textit{bonafide} and \textit{spoof} label-sequences, $y \in \mathit{\{bonafide,spoof\}}$ is the ground-truth class, and $w_y$ compensates for class imbalance.
To further promote separation between competing hypotheses, we optionally incorporate a margin-based regularization term. We define:
\begin{equation}
d =
\begin{cases}
\ell_{\mathrm{sp}} - \ell_{\mathrm{bf}}, & y = \mathrm{spoof} \\
\ell_{\mathrm{bf}} - \ell_{\mathrm{sp}}, & y = \mathrm{bonafide}
\end{cases}
\end{equation}
and handle cases where the difference becomes smaller than a predefined margin:
\begin{equation}
\mathcal{L}_{\mathrm{margin}} =
\begin{cases}
\max(0, m_{\mathrm{sp}} - d), & y = \mathrm{spoof} \\
\max(0, m_{\mathrm{bf}} - d), & y = \mathrm{bonafide}
\end{cases}
\end{equation}
where $m_{\mathrm{sp}}$ and $m_{\mathrm{bf}}$ are fixed margin hyper-parameters. The final loss is defined as:
\begin{equation}
\mathcal{L}_{\mathrm{total}} =
\mathcal{L}_{\mathrm{CE}} + \lambda_{\mathrm{margin}} \mathcal{L}_{\mathrm{margin}}
\end{equation}

\section{Databases and Evaluation}
\label{sec:databases_Evaluation_Metrics}
This section describes the databases and evaluation metric used in this work.
\subsection{ASVspoof2019 LA Database}
The ASVspoof2019 \textit{logical access} (LA) database provides spoofed and bonafide utterances generated under clean recording conditions, as summarized in~\autoref{tab:ASVSpoofDATA}. 
The training (Train) and development (Dev.) sets of the LA partition include the VC and TTS attacks (A01-A06), while the evaluation (Eval.) set additionally introduces a combination of VC and TTS attacks. The evaluation set includes 11 unseen spoofing attacks (A07-A15, A17, A18), along with two overlapping attacks (A16, A19) produced using algorithms seen in the training set but trained on different databases~\cite{wang2020asvspoof}. 
\begin{table}[t]
  \centering
  \caption{Number of bonafide and spoofed utterances in ASVspoof2019 LA, ASVspoof2021 LA, ASVspoof2021 DF, and ASVspoof5 databases.}
  \label{tab:ASVSpoofDATA}
  \begin{tabular}{|c|c|c|c|}
  \hline
  \textbf{Database} & \textbf{Set} & \textbf{bonafide} & \textbf{Spoof} \\
  \hline
  \multirow{3}{*}{ASVspoof2019 LA} 
    & Train  & 2,580   & 22,800 \\
  \cline{2-4}
    & Dev.   & 2,548   & 22,296 \\
  \cline{2-4}
    & Eval.  & 7,355   & 63,882 \\
  \hline

  \multirow{1}{*}{ASVspoof2021 LA} 
    & Eval.  & 14,816   & 133,360 \\
  \hline

  \multirow{1}{*}{ASVspoof2021 DF} 
    & Eval.  & 14,869 & 519,059 \\
  \hline

  \multirow{3}{*}{ASVspoof5} 
    & Train  & 18,797   & 163,560 \\
  \cline{2-4}
    & Dev.   & 31,334   & 109,616 \\
  \cline{2-4}
    & Eval.  & 138,688  & 542,086 \\
  \hline
  \end{tabular}
  \vspace{-4mm}
\end{table}

\subsection{ASVspoof 2021 LA and DF Databases}
The ASVspoof2021 database~\cite{10155166} provides an evaluation-only set and splits to LA and DF partitions (as shown in~\autoref{tab:ASVSpoofDATA}). Therefore, the participants are required to rely on the training and development partitions of the ASVspoof2019 LA database~\cite{wang2020asvspoof}. 
In the ASVspoof2021 LA task, speech signals are processed through telephony codecs (e.g., OPUS, G.722), while in the ASVspoof2021 DF task, speech signals are processed through media compression codecs (e.g., MP3, M4A, OGG).

\subsection{ASVspoof5 Database}
The ASVspoof5 database follows a similar three-part structure, as shown in~\autoref{tab:ASVSpoofDATA}. The training set includes eight TTS attacks (A01-A08), the development set contains eight different attacks (A09-A16). The evaluation set includes 16 unseen spoofing attacks (A17-A32), including adversarial attacks, while also applying transmission and compression codecs, including neural codecs, to both bonafide and spoofed speech~\cite{wang24_asvspoof,WANG2026101825}. 

The performance is assessed using the EER metric, computed from the log-likelihood difference for the Spooftral CM system and from the classifier scores for linear probing and the Spooftral-Enc CM system.

\section{Experiments and Results}
\label{sec:Experiments_and_Results}
This section presents the experiments and results. All experiments are based on the Voxtral-mini-3B model~\cite{liu2025voxtral}, which is based on the Ministral-3B model~\cite{liu2026ministral}, and uses \textit{brain floating point} (bfloat16) precision~\cite{kalamkar2019study}. The \textit{confidence intervals} (CIs) were computed using 1,000 bootstrap iterations with a significance level of $5\%$~\cite{Confidence_Intervals}.

The embeddings are extracted from the last hidden layers of the Voxtral LLM decoder and audio adapter, as illustrated in~\autoref{fig:architecture}.
Inspired by~\cite{serrano2025improving}, the transformer output is denoted as $\mathbf{Z} \in \mathbb{R}^{T \times F}$, where $T$ denotes the number of temporal frames and $F$ denotes the feature dimension. Mean pooling over the temporal dimension produces a fixed-dimensional embedding $\mathbf{e} \in \mathbb{R}^{F}$, which is then fed into a linear layer for binary spoofing detection, with all preceding layers frozen and only the linear layer is trained (linear probing approach as shown in~\cite{zaiem23b_interspeech}).

Since the pre-trained Voxtral model is not optimized for spoofing detection, it may generate different textual responses corresponding to the same spoofing decision (e.g., spoof, fake, or synthetic), making EER evaluation based on text generation unreliable. Therefore, we use linear probing on frozen representations to directly assess the spoof-discriminative information encoded at different stages of the Voxtral model.

The results obtained using a fixed prompt \textbf{``Is this speech spoof or bonafide?''}  are shown in~\autoref{tab:linear_probing_combined} (each Dev. and Eval. pair uses the same trained linear layer).
Based on~\autoref{tab:linear_probing_combined}, performance obtained using embeddings from the LLM decoder is inferior to that obtained using the audio-adapter representations, which are derived from the Whisper-based audio encoder~\cite{Radford2022RobustSR}. These results suggest that spoofing detection requires task-specific training, as performance consistently degrades after representations are passed through the LLM decoder. A reasonable explanation is that Voxtral is pre-trained primarily for semantic tasks, such as ASR and \textit{speech-to-text translation}, as shown in~\cite{liu2025voxtral}, rather than for acoustic tasks such as spoofing detection.
\begin{figure}
    \centering
    \includegraphics[width=1\linewidth]{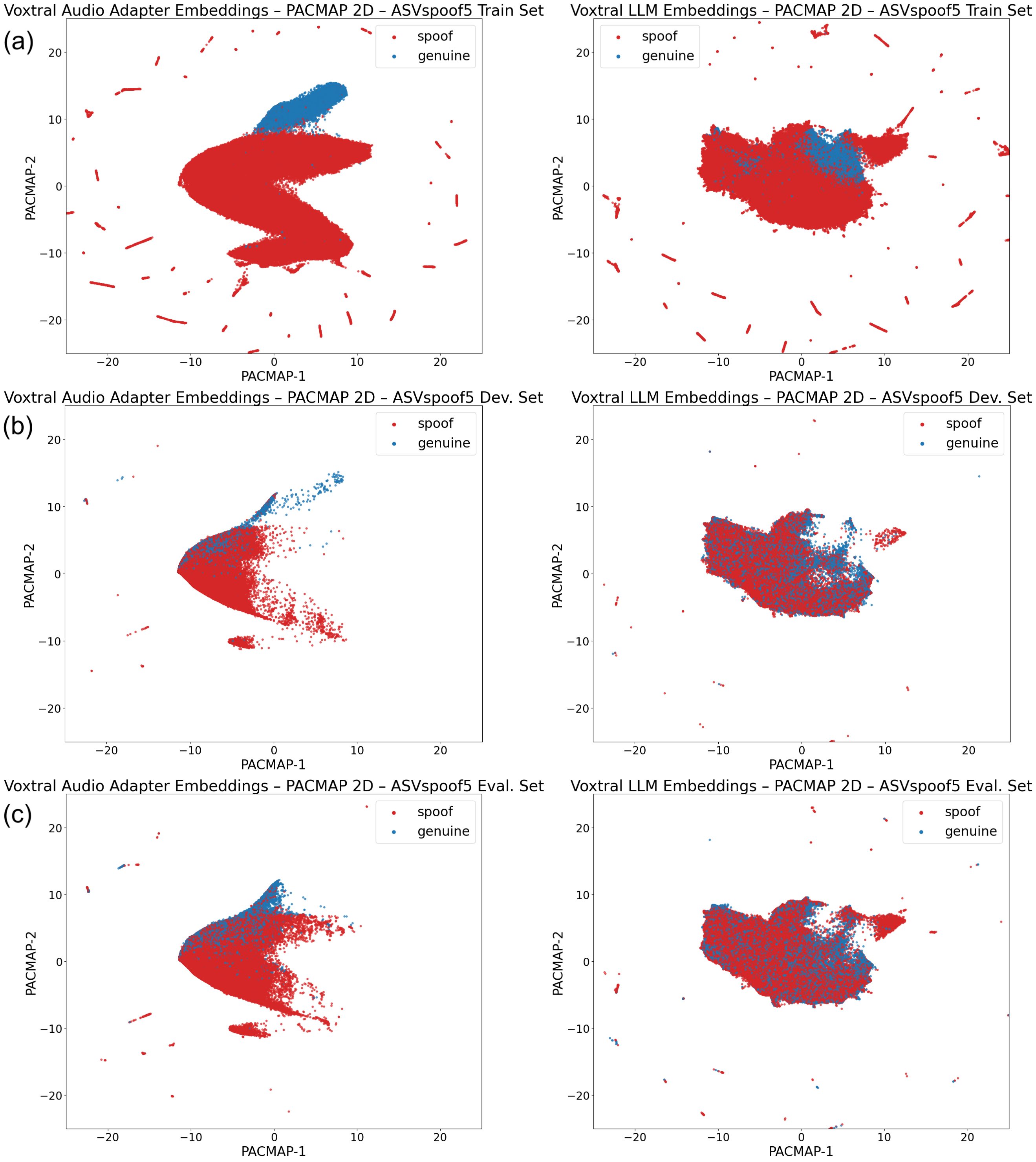}
    \caption{PaCMAP projections of audio adapter (left) and LLM (right) embeddings across ASVspoof5 sets: (a) training set, (b) Dev. set, (c) Eval. set.}
    \label{fig:PaCMAP_asvspoof5}
\end{figure}

To visualize embedding structure after the audio adapter and the LLM layers, we project the embeddings into two dimensions using the PaCMAP dimensionality reduction algorithm~\cite{JMLR_PaCMAP}, which is trained on the ASVspoof5 training set and then applied to the development and evaluation sets.

As shown in~\autoref{fig:PaCMAP_asvspoof5}, embeddings extracted after the audio adapter show clear separation between bonafide and spoofed samples on the ASVspoof5 training set. This is consistent with~\cite{liu2025voxtral}, as Voxtral was not trained for the spoofing detection task, but was pre-trained on an audio-text corpus that includes TTS data, while the ASVspoof5 training set consists exclusively of TTS-based attacks, which may explain the strong separation observed on the ASVspoof5 training set.
In contrast, embeddings obtained after the LLM decoder show reduced separation, indicating that the LLM layers, which are optimized for semantic processing, attenuate spoofing-related cues and degrade class separability.

A similar behavior is observed for the development and evaluation sets, where the embeddings after the audio adapter show better separability. 

\begin{table}[t]
\centering
\caption{Linear probing performance after the audio adapter and LLM layers on ASVspoof databases (EER~\%). 
CI is shown below each value.}
\label{tab:linear_probing_combined}
\setlength{\tabcolsep}{2pt}

\begin{tabular}{|c|c|c|c|c|}
\hline
\textbf{Database} &
\multicolumn{2}{c|}{\textbf{LLM}} &
\multicolumn{2}{c|}{\textbf{Audio Adapter}} \\
\cline{2-5}
 & \textbf{Dev.} & \textbf{Eval.} & \textbf{Dev.} & \textbf{Eval.} \\
\hline

ASVspoof2019 LA
 & \makecell{6.60 \\ {\scriptsize (5.96,7.09)}} 
 & \makecell{7.17 \\ {\scriptsize (6.91,7.55)}}
 & \textbf{\makecell{2.79 \\ {\scriptsize (2.37,3.07)}}} 
 & \textbf{\makecell{5.48 \\ {\scriptsize (5.20,5.73)}}}  \\
\hline

ASVspoof2021 LA
 & \makecell{9.26 \\ {\scriptsize (8.74,9.84)}} 
 & \makecell{17.34 \\ {\scriptsize (16.97,17.70)}}
 & \textbf{\makecell{2.79 \\ {\scriptsize (2.37,3.07)}}} 
 & \textbf{\makecell{13.57 \\ {\scriptsize (13.30,13.87)}}} \\
\hline

ASVspoof2021 DF
 & \makecell{7.02 \\ {\scriptsize (6.40,7.50)}} 
 & \makecell{19.20 \\ {\scriptsize (18.96,19.52)}}
 & \textbf{\makecell{2.79 \\ {\scriptsize (2.37,3.07)}}} 
 & \textbf{\makecell{9.64 \\ {\scriptsize (9.49,9.86)}}} \\
\hline

ASVspoof5
 & \makecell{20.73 \\ {\scriptsize (20.49,20.96)}} 
 & \makecell{19.09 \\ {\scriptsize (18.98,19.19)}}
 & \textbf{\makecell{11.58 \\ {\scriptsize (11.40,11.74)}}} 
 & \textbf{\makecell{9.75 \\ {\scriptsize (9.68,9.83)}}} \\
\hline

\end{tabular}
\vspace{-4mm}
\end{table}

Next, we trained two variants of the Spooftral model using DoRA~\cite{liu2024dora}, which extends \textit{low-rank adaptation} (LoRA)~\cite{hu2022lora} by decomposing the weights of the adapted layers into magnitude and direction components.
The full ALM model, denoted as \textit{Spooftral} (Voxtral + DoRA), adapts Voxtral for the spoofing detection task using the architecture shown in~\autoref{fig:architecture}.
In addition, we trained an audio encoder-and-adapter-only model, denoted \textit{Spooftral-Enc}, which excludes the LLM decoder and consists of a Whisper-based audio encoder with DoRA applied to the attention layers, followed by an audio adapter and a lightweight trainable classification head. The classification head comprises a linear projection from 3072 to 256 dimensions, a GELU activation, a dropout layer with probability of $0.1$, and a final linear layer from 256 to 2 output dimensions.

In both models, DoRA was applied only to the attention projection layers (query, key, value, and output) with a rank of 16 and $\alpha=32$. For the ASVspoof5 database, the DoRA rank increased to 32 and $\alpha$ to 64; this configuration is denoted as \textit{Attention DoRA} and is used in~\autoref{tab:experiments_spooftral}. The audio adapter layers remained fully trainable.
The training process includes data augmentation with additive noise and codec perturbations (e.g., MP3 and OPUS). To avoid transforming bonafide speech utterances into speech that resembles spoofed utterances, a maximum of three augmentations in parallel were applied to each utterance.
We used the AdamW optimizer with a cosine learning-rate scheduler and a learning rate of $10^{-4}$, with gradient accumulation of 8, training for up to 20 epochs with a batch size of 32 (before gradient accumulation). Spooftral-Enc was trained with the cross-entropy loss, whereas Spooftral was trained using the proposed generative label-likelihood classification method. In the Spooftral variant the margin contribution was controlled by $\lambda_{\mathrm{margin}}$, with $m_{\mathrm{bf}}=0.1$, $m_{\mathrm{sp}}=0.05$, and $\lambda_{\mathrm{margin}}=0.05$ in all experiments using margin regularization.

To address the class imbalance in the training sets, class weights were computed using the square-root inverse frequency, $w_i = \sqrt{\frac{N}{n_i}}$, where $n_i$ is the number of utterances in class $i$ and $N$ is the total number of utterances. The resulting weights were normalized such that their sum equals 2. For all ASVspoof databases, the normalized class weights were approximately $(w_{\mathrm{bf}}, w_{\mathrm{sp}})=(1.5,0.5)$.

The results obtained with Spooftral and Spooftral-Enc on the ASVspoof databases are shown in~\autoref{tab:experiments_spooftral}.
\begin{table}[t]
\centering
\caption{Performance comparison of Spooftral and Spooftral-Enc across the ASVspoof databases (EER~\%). CI is shown below each value.}
\label{tab:experiments_spooftral}
\renewcommand{\arraystretch}{1}
\begin{tabular}{|l|c|c|c|c|}
\hline
\multicolumn{1}{|c|}{\textbf{Database}} &
\multicolumn{2}{c|}{\textbf{Spooftral}} &
\multicolumn{2}{c|}{\textbf{Spooftral-Enc}} \\
\cline{2-5}
& \textbf{Dev.} & \textbf{Eval.} & \textbf{Dev.} & \textbf{Eval.} \\
\hline

ASVspoof2019 LA
& \textbf{\makecell{0.02 \\ {\scriptsize (0.00,0.11)}}}
& \makecell{0.41 \\ {\scriptsize (0.35,0.47)}}
& \makecell{0.03 \\ {\scriptsize (0.00,0.11)}}
& \textbf{\makecell{0.40 \\ {\scriptsize (0.34,0.50)}}}
\\
\hline

ASVspoof2021 LA
& \textbf{\makecell{0.02 \\ {\scriptsize (0.00,0.11)}}}
& \textbf{\makecell{6.46 \\ {\scriptsize (6.25,6.69)}}}
& \makecell{0.03 \\ {\scriptsize (0.00,0.12)}}
& \makecell{ 6.70 \\ {\scriptsize (6.43,6.89)}}
\\
\hline

ASVspoof2021 DF
& \textbf{\makecell{0.05 \\ {\scriptsize (0.00,0.09)}}}
& \textbf{\makecell{3.88 \\ {\scriptsize (3.78,3.97)}}}
& \makecell{0.05 \\ {\scriptsize (0.00,0.11)}} 
& \makecell{4.76 \\ {\scriptsize (4.61,4.91)}}
\\
\hline

ASVspoof5
& \textbf{\makecell{4.50 \\ {\scriptsize (4.41,4.61)}}} 
& \textbf{\makecell{4.25 \\ {\scriptsize (4.20,4.31)}}}
& \makecell{5.29 \\ {\scriptsize (5.18,5.41)}}
& \makecell{5.08 \\ {\scriptsize (5.03,5.14)}} \\
\hline

\end{tabular}
\vspace{-4mm}
\end{table}
Spooftral and Spooftral-Enc achieve strong performance across the ASVspoof databases. The observed performance differences across the evaluated databases may be explained by several factors. First, ASVspoof5 provides a larger and more diverse training set, allowing DoRA to better adapt to spoofing-specific artifacts while preserving the pre-trained model weights. Second, the Whisper-based audio encoder in Voxtral is primarily optimized for robust speech transcription and therefore learns representations that are relatively invariant to codec and channel distortions. While this property is beneficial for automatic speech recognition, it may reduce sensitivity to spoofing-related artifacts and consequently limit generalization under the mismatched conditions.

Another important difference lies in the database design. In ASVspoof2021, the training and development sets are based on ASVspoof2019 and therefore contain the same attack types, whereas the evaluation set consists of unseen attacks. In contrast, the ASVspoof5 training and development sets contain different attack types, encouraging the model to learn more generalizable spoofing representations and leading to an operating point that transfers better to unseen attacks.

Nevertheless, the EERs obtained on the development and evaluation sets in the ASVspoof5 database, despite their different spoofing attacks, suggest good generalization across different attack types.

Furthermore, Spooftral consistently outperforms Spooftral-Enc on most evaluation sets, except on the ASVspoof2019 LA evaluation set, where the performance difference is minor. This result indicates that although the frozen LLM representations alone are not optimal for spoof detection, task-specific adaptation enables the LLM to learn spoof-discriminative representations that complement the acoustic encoder. Finally, both Spooftral variants outperform the Whisper-based spoofing detection approach reported in~\cite{10.1007/978-3-031-61382-1_8}, demonstrating the effectiveness of the model for the spoofing detection task.

To further analyze the generative label-likelihood classification method, we trained two Spooftral models using different instruction interfaces. The two interfaces differ only in the prompt and the target label sequences used during training and inference, while the DoRA configuration and all other training settings remain identical. The results are summarized in~\autoref{tab:instruction_interface}.

\begin{table}[t]
\centering
\caption{Instruction interface ablation for Spooftral on ASVspoof5 without margin regularization (EER~\%). CI below each
value.}
\label{tab:instruction_interface}
\begin{tabular}{|l|c|c|}
\hline
\textbf{Instruction Interface} & \textbf{Dev. EER} & \textbf{Eval. EER} \\
\hline
\makecell[l]{Prompt: ``Classify the audio as real or fake.''\\
Labels: \textit{real} / \textit{fake}}
& \makecell{5.67 \\ {\scriptsize (5.56,5.82)}} & \makecell{4.47 \\ {\scriptsize (4.42,4.52)}} \\
\hline
\makecell[l]{Prompt: ``Is this speech spoof or bonafide?''\\
Labels: \textit{bonafide} / \textit{spoof}}
& \textbf{\makecell{4.77 \\ {\scriptsize (4.66,4.89)}}} & \textbf{\makecell{4.46 \\ {\scriptsize (4.39,4.53)}}} \\

\hline
\end{tabular}
\end{table}

The \textit{bonafide/spoof} instruction interface achieved lower EERs than the \textit{real/fake} interface, with a more pronounced improvement on the development set. This observation suggests that instruction design may influence performance in generative label-likelihood classification, potentially due to differences in how label sequences interact with the pre-trained ALM representations. However, further studies are required to disentangle the independent effects of the prompt and target label sequences on these representations. Therefore, the \textit{bonafide/spoof} instruction interface was used in the other experiments.

\begin{table}[t]
\centering
\caption{Comparison of Spooftral and Spooftral-Enc versus other SSL-based CMs on ASVspoof5 (EER~\%). CI below each value.}
\label{tab:comparison_asvspoof5}
\begin{tabular}{|l|c|c|}
\hline
\textbf{System} & \textbf{Dev.} & \textbf{Eval.} \\
\hline
S10 Fusion$^\dagger$~\cite{kulkarni24_asvspoof} & -- & 11.24 \\
Fusion of WavLM-ResNet18-SA$^\dagger$$^\ast$~~\cite{chan24_asvspoof} & 0.64 & 7.01  \\
SSL-IVSPT$^\ast$~\cite{guo24_asvspoof} & 0.76 & 5.99 \\
SLIM$^\dagger$$^\ast$~\cite{zhu24_asvspoof} & -- & 5.50 \\
Best open-condition submission$^\dagger$$^\ast$~\cite{chen24_asvspoof} & -- & 2.59 \\
\hline
Spooftral (LLM Attention DoRA) 
& \makecell{7.89 \\ {\scriptsize (7.75,8.03)}} 
& \makecell{7.01 \\ {\scriptsize (6.94,7.08)}} \\

Spooftral (Attention DoRA) 
& \makecell{4.77 \\ {\scriptsize (4.66,4.89)}} 
& \makecell{4.46 \\ {\scriptsize (4.39,4.53)}} \\

\textbf{Spooftral (Attention DoRA + Margin)} 
& \textbf{\makecell{4.50 \\ {\scriptsize (4.41,4.61)}}} 
& \textbf{\makecell{4.25 \\ {\scriptsize (4.20,4.31)}}} \\

Spooftral-Enc (Attention DoRA)
& \makecell{5.29 \\ {\scriptsize (5.18,5.41)}}
& \makecell{5.08 \\ {\scriptsize (5.03,5.14)}} \\
\hline
\end{tabular}
\begin{flushleft}
{\footnotesize 
$\ast$ uses additional training database\\
$\dagger$ uses fusion
}
\end{flushleft}
\vspace{-4mm}
\end{table}

\autoref{tab:comparison_asvspoof5} compares different Spooftral training framework configurations on the ASVspoof5 database. The first variant, denoted \textit{Spooftral (LLM Attention DoRA)}, applies DoRA only to the LLM attention layers while the audio encoder is frozen and the audio adapter is trainable. The second variant, denoted \textit{Spooftral (Attention DoRA)}, applies DoRA to all attention projection layers, including those of the audio encoder. The third variant, denoted \textit{Spooftral (Attention DoRA + Margin)}, further incorporates the proposed margin regularization term. The obtained results show that the proposed margin regularization consistently improves the Attention DoRA configuration, yielding lower EERs on both the development and evaluation sets of the ASVspoof5 database.

Among them, SLIM~\cite{zhu24_asvspoof}, SSL-IVSPT~\cite{guo24_asvspoof}, and the USTC-KXDIGIT system (best open-condition ASVspoof5 challenge submission)~\cite{chen24_asvspoof} achieve strong performance, often relying on audio encoder fusion, additional spoofing-detection training data, or both.

In contrast, our Spooftral variants, built on the pre-trained Voxtral-mini model~\cite{liu2025voxtral}, achieve competitive performance using lightweight adaptation alone. The Spooftral model (DoRA rank 32, $\alpha=64$) requires only $\sim$55.1M trainable parameters, corresponding to just $1.17\%$ of its 4.71B total parameters. Similarly, Spooftral-Enc uses only $\sim$37.4M trainable parameters ($5.55\%$ of its 674M total parameters). Both models achieve these results without model fusion or additional spoofing-specific training databases.

\begin{figure}[t]
    \centering
    \includegraphics[width=1\linewidth]{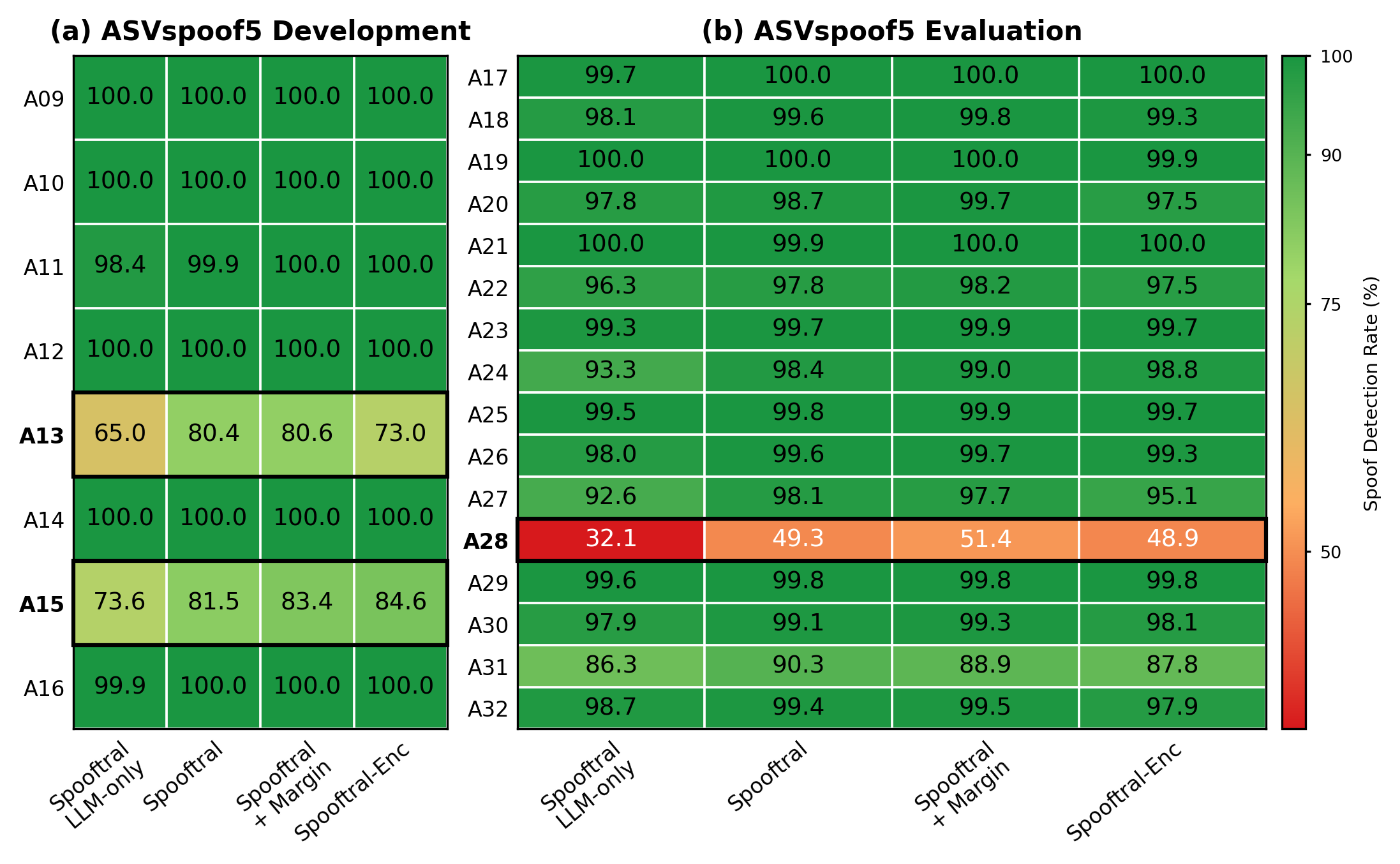}
    \caption{Per-attack spoof detection rate (\%) on the ASVspoof5 development and evaluation sets using the EER threshold of each set.}
    \label{fig:asvspoof5_attack_heatmap}
    \vspace{-4mm}
\end{figure}

For a deeper analysis of the proposed models, we examine their performance across individual spoofing attacks.
As shown in~\autoref{fig:asvspoof5_attack_heatmap}, the per-attack spoof detection rate based on the EER threshold of each set shows a consistent pattern across all evaluated Spooftral and Spooftral-Enc models trained on the ASVspoof5 database. On the development set, the A13 (VC-based StarGANv2-VC~\cite{li2021starganv2}) and A15 (VAE-GAN~\cite{albadawy2020voice}) attacks consistently exhibit the lowest spoof detection rate. In contrast, on the evaluation set, the A28 (pre-trained YourTTS~\cite{casanova2022yourtts}) attack consistently yields a lower spoof detection rate than the remaining attacks. This observation is consistent with other analyses~\cite{schafer24_asvspoof,WANG2026101825,martindonas24_asvspoof}, which also identified the A28 attack as one of the most challenging attacks for CM systems~\cite{schafer24_asvspoof}.

To the best of our knowledge, no previous ALM-based CM has been reported on the ASVspoof5 database. Although the linear probing results indicate stronger separability in the audio adapter representations, the superior performance of Spooftral demonstrates that the LLM component can contribute effectively to spoof detection after task-specific DoRA adaptation. These findings motivate further task-aware adaptation and architectural modifications for ALM-based spoofing detection systems.

\section{Discussion and Conclusions}
\label{sec:Conclusions_and_Discussion}
This work investigated the Voxtral ALM framework for spoofing detection and examined whether an instruction-guided Voxtral model can contribute to spoofing detection beyond acoustic classifiers while serving as an initial step toward integrating CM capabilities into a unified ALM framework. We reformulate spoofing detection as an instruction-guided generative label-likelihood classification task, where bonafide and spoof hypotheses are scored using length-normalized label-sequence likelihoods. We analyze how spoof-discriminative information propagates across model stages by comparing representations before and after the LLM decoder. 
The results show that spoof cues are more separable before the LLM decoder in the frozen setting, consistent with an objective mismatch whereby the decoder, optimized for semantic robustness, reduces the separability of fine-grained acoustic artifacts critical for spoof detection. 
However, the superior performance of Spooftral after DoRA adaptation demonstrates that the LLM decoder component can effectively contribute to spoof detection when adapted to the task. Furthermore, the larger scale and greater diversity of the ASVspoof5 training set may enable the Spooftral architecture to learn more robust spoof-discriminative representations, leading to better generalization and a competitive EER of 4.25\% on the ASVspoof5 evaluation set compared with the submissions to the ASVspoof5 Challenge~\cite{wang24_asvspoof}.

Future work will extend this framework to jointly address spoofing detection and other speech tasks within a unified ALM framework. We plan to investigate audio encoders that are more sensitive to acoustic artifacts and alternative speech representations, as current ALM frameworks are based on the Whisper-based audio encoder~\cite{tang2023salmonn,Qwen-Audio,ghosh2025audio,kumar2025falcon3audio}. We also plan to investigate alternative adaptation and training strategies~\cite{refael2025adarankgrad,meng2024pissa,3692070.3692782} and incorporate reasoning-based training strategies in the unified ALM framework, which may further improve performance on the spoofing detection task and enhance generalization to previously unseen spoofing attacks.


\bibliographystyle{IEEEtran}
\bibliography{references}
\end{document}